\documentclass[aps,twocolumn,superscriptaddress,altaffilletter,lengthcheck,
tightenlines,showpacs,showkeys]{revtex4}

\newcommand{\al}{\alpha}
\newcommand{\bb}{\beta}
\newcommand{\D}{\Delta}
\newcommand{\ben}{\begin{eqnarray}}
\newcommand{\een}{\end{eqnarray}}
\newcommand{\be}{\begin{equation}}
\newcommand{\ee}{\end{equation}}
\newcommand{\ba}{\begin{eqnarray}}
\newcommand{\ea}{\end{eqnarray}}
\newcommand{\n}{\label}

\newcommand{\ga}{\gamma}
\newcommand{\ro}{\rho}

\newcommand{\bn}{\begin{equation}\label}
\newcommand{\m}{MHRDE }

\usepackage[dvipdf]{epsfig}
\usepackage{color}

\begin{document}

\title{ Dark matter and  Ricci-like holographic dark energy \\
 coupled through a quadratic interaction}

\author{Luis P. Chimento}\email{chimento@df.uba.ar}
\affiliation{Departamento de F\'{\i}sica, Facultad de Ciencias Exactas y Naturales,  Universidad de Buenos Aires, Ciudad Universitaria, Pabell\'on I, 1428 Buenos Aires, Argentina}
\author{\bfseries{Mart\'{\i}n G. Richarte}}\email{martin@df.uba.ar}
\affiliation{Departamento de F\'{\i}sica, Facultad de Ciencias Exactas y Naturales,  Universidad de Buenos Aires, Ciudad Universitaria, Pabell\'on I, 1428 Buenos Aires, Argentina}

\date{\today}
\bibliographystyle{plain}


\begin{abstract}
We  examine a spatially flat Friedmann-Robertson-Walker (FRW) universe filled with interacting dark matter and a modified holographic Ricci dark energy (MHRDE). The interaction term is selected as a significant rational function of the total energy density and its first derivative homogeneous of degree. We show that the effective one-fluid obeys the equation of state of a relaxed Chaplygin gas, then the universe  turns to be dominated by pressureless dark matter at early times and undergoes an accelerated expansion in the far future driven by a strong negative pressure. Performing a $\chi^{2}$-statistical analysis  with the observational Hubble data and the Union2 compilation of SNe Ia, we  place some constraints on cosmological parameters analyzing the feasibleness of the modified holographic Ricci ansatz. It turned that MHRDE gets the accelerated expansion faster than the $\Lambda$CDM model. Finally, a new model with a component that does not exchange energy with the interacting dark sector is presented for studying bounds on the dark energy at early times.

\end{abstract} 
\vskip 1cm


\bibliographystyle{plain}

\maketitle


\section{Introduction}
The existence of a mysterious fuel called dark energy (DE) stems from astronomical observations which indicate that the Universe is currently undergoing an accelerating phase driven by an exotic component. This tremendous fact has been confirmed by a plethora of observational  tests such as high redshift Hubble diagram of type Ia supernovae as standard candles \cite{c1a} and  accurate measurements of  cosmic microwave background (CMB) anisotropies\cite{c4a}. Despite DE represents more than $70\%$ of the total energy of the Universe, the current dark energy density is about 120 order the magnitude smaller than the energy scales at the end of inflation, so one of the main challenge in the modern cosmology is to understand this missmacth. One way to alleviate the aforesaid problem is working within the context of dynamical dark energy models, leaving aside the standard $\Lambda$CDM model. Besides, the necessity of a dark matter component comes from astrophysical evidences of colliding galaxies, gravitational lensing of mass distribution or power spectrum  of clustered matter \cite{DMobserva}, \cite{dme}. Moreover, the astrophysical observations from the galactic to the cosmological scales sustain that dark matter represents nearly $25\%$ of the total energy-matter of the Universe; this  substantial unvisible and non-baryonic component is the major agent responsible for the large-structure formation in the Universe\cite{DMobserva}. 

Another point of debate refers to the coincidence problem, namely,  why  dark energy and dark matter have energy densities of the same orders of magnitude despite the fact that both densities dilutes at different rates?.  Motived to understand both problems  one could consider an exchange of energy between the dark components, i.e., the dark matter not only can feel the presence of  the dark energy through a  gravitational expansion of the Universe but also  can interact between them \cite{jefe1}.  More precisely,  a coupling between dark energy and dark matter changes the background evolution of the dark sector allowing us to constrain a particular type of interaction and giving us the possibility of studying the coincidence within the context of interacting dark sector also.

The aim of present brief article is to examine the exchange of energy between  dark matter and the dark energy being the interaction selected  as a linear combination of the total dark energy density, its derivative plus a non-linear term. The DE is associated with a modified holographic Ricci ansatz. We  explicitly show that the effective one-fluid obeys the equation of state of a relaxed Chaplygin gas. Later on, applying a $\chi^{2}$-statistical method to the  Hubble data and  the Union2 compilation of SNe Ia some constraints are placed on the cosmological parameters. Using their best--fit values, we confront our model with the standard $\Lambda$CDM. Besides, the issue of dark energy at early times is also discussed when a third component is added. At the end, we summarize  our findings. 

\section{The model }
Considering the effective  quantum field theory  it has been shown that  the zero-point  energy  of a system  with size $L$ should no exceed  the mass  of a black hole with the same size, thus $L^{3}\ro_{\Lambda}\leq LM^{2}_{P}$, where $\ro_{\Lambda}$ corresponds to the quantum zero-point energy density \cite{holo1} with $M^{-2}_{P}= 8\pi G$. This relation gives a link between  the ultraviolet cutoff, define through $\ro_{\Lambda}$, and the infrared cutoff  which is encoded by the scale $L$. Taking into account this novel principle within the cosmological context, one  assumees that the dark energy density of the universe $\ro_{x}$ takes the same form of the vacuum energy, thus $\ro_{\Lambda}=\ro_{x}$. Using the largest $L$ as the one saturating the above inequality, it turns out to be the holographic dark energy is given by $\rho_{x}=3c^{2}M^{2}_{~P}L^{-2}$ with $c$ a numerical factor \cite{hde1}, \cite{odi1}. Many different proposals for the cutoff $L$ have been studied in the literature \cite{odi2}, \cite{go2a}, \cite{go2b}, \cite{HLnew}, \cite{HL1}, \cite{HoloG}.

Our starting point is to consider an holographic cosmological model with an IR cutoff  given by the modified Ricci's radius so that  $L^{-2}$ is  a linear combination of $\dot H$ and $H^2$\cite{go}, \cite{HL2}, \cite{HL3}. After that, the modified holographic Ricci dark energy (MHRDE) becomes 
\be
\n{03}
\ro_x=\frac{2}{\al -\bb}\left(\dot H + \frac{3\al}{2} H^2\right).
\ee
Here $H = \dot a/a$ is the Hubble expansion rate, $a$ is the scale factor and $\al, \bb$ are free constants. Introducing the variable $\eta = \ln(a/a_0)^{3}$, with $a_0$ the present value of the scale factor and $' \equiv d/d\eta$, the above MHRDE (\ref{03}) becomes a modified conservation equation (MCE) for the cold dark matter $\ro_c$ and the MHRDE
\be
\n{06}
\ro'=-\al\ro_c -\bb\ro_x,
\ee
after using the Friedmann equation, 
\be
\n{00}
3H^2= \ro_c + \ro_x,
\ee 
for a spatially flat FRW cosmology and $\ro=\ro_c + \ro_x$. The MCE (\ref{06}) looks as it were a conservation equation for both dark components with constant equations of state. In connection with observations on the large scale structures, which seems to indicate that the Universe  must have been dominated by nearly pressureless components, we assume that $\ro_c$ includes all these components and has an equation of state $p_c=0$ while the \m has a barotropic index $\omega_x=p_x/ \ro_x$, so that the whole conservation equation (WCE) becomes 
\be
\n{05}
\ro'=-\ro_c -(\omega_x +1)\ro_x.
\ee
The compatibility between the MCE (\ref{06}) and the WCE (\ref{05}) yields a linear dependence of the equation of state of the MHRDE 
\be
\n{07}
\omega_x=(\al -1)r+ (\beta-1),
\ee
with the ratio of both dark components $r=\ro_c/\ro_x$. Solving the linear algebraic system of equations (\ref{06}) and $\ro=\ro_c+\ro_x$ we obtain both dark energy densities as functions of $\ro$ and $\ro'$
\be
\n{10}
\ro_c= - \frac{\bb \ro +\ro '}{\D\ga}, \qquad \ro_x= \frac{\al \ro +\ro '}{\D\ga},
\ee
with $\D\ga = \al -\bb$, while the total pressure is  $p_{x} =-\ro -\ro'$. From now on we will use the MCE (\ref{05}) instead of the WCE with variable $\omega_{x}$ because it is simpler, and introduce an interaction between both dark components through the term $3HQ$ into the MCE (\ref{05}) with constant coefficients, so 
\be
\n{08}
\ro_c' + \al \ro_c = - Q,
\ee
\be
\n{09}
\ro_x' + \bb \ro_x = Q.
\ee
Finally, from Eqs. (\ref{10}) and (\ref{08}), we obtain the source equation \cite{jefe1} for the energy density $\ro$
\be
\n{14}
\ro''+(\al + \bb)\ro' + \al\bb\ro =  Q\D\ga.
\ee
 
Now we consider cosmological models where the interaction $Q$ between both dark components is nonlinear and includes a set of terms  which are homogeneous of degree 1 in the total energy density and its first derivative \cite{jefe1}, 
\be
\n{Q}
Q=\frac{(\al\beta -1)}{\Delta\gamma}\,\ro+\frac{(\al + \beta -\nu-2)}{\Delta\gamma}\,\ro'-\frac{\nu\ro'^{2}}{\ro\Delta\gamma},
\ee
where $\nu$ is a positive constant that parameterizes the interaction term $Q$. Replacing (\ref{Q}) into (\ref{14}) it turns into a nonlinear second order differential equation for the energy density: $\ro\ro''+(2+\nu)\ro\ro'+\nu\ro'^{2}+\ro^2=0$. Introducing the new variable $y=\ro^{(1+\nu)}$ into the latter equation one gets a second order linear differential equation $y''+(2+\nu)y'+(1+\nu)y=0$, whose solutions allow us to write the energy density as
\be
\label{Et}
\ro=\left[\ro_{10}a^{-3}+\ro_{20}a^{-3 (1+\nu)}\right]^{1/(1+\nu)}
\ee
being $\ro_{10}$ and $\ro_{20}$ positive constants. From Eqs. (\ref{10})-(\ref{Et})  and using that  $p =-\ro -\ro'$, we have both dark energy densities and the total pressure
\be
\n{cI}
\ro_c=\frac{-\ro}{\al-\bb}\left[\bb-1+\frac{\nu}{(1+\nu)(1+\ro_{20}a^{-3\nu}/\ro_{10})}\right],\,\,\,
\ee
\be
\n{xI}
\ro_x=\frac{\ro}{\al-\bb}\left[\al-1+\frac{\nu}{(1+\nu)(1+\ro_{20}a^{-3\nu}/\ro_{10})}\right],\,\,\,\,\,\,\,\,\,\,\,\,	
\ee
\be
\n{24}
p=-\frac{\nu\ro_{10}}{1+\nu}\,\frac{a^{-3}}{\ro^\nu}.
\ee
From these equations we see that an initial model of interacting dark matter and dark energy can be associated with 
an effective one-fluid description of an unified cosmological scenario where the effective one-fluid, with energy density $\ro=\ro_c+\ro_x$ and pressure (\ref{24}), obeys the equation of state of a relaxed Chaplygin gas $p=b\ro+f(a)/\ro^\nu$, where $b$ is a constant \cite{jefe1}. The effective barotropic index $\omega=p/\ro=\omega_x\ro_x/\ro$ reads,
\be
\n{eb}
\omega=-\frac{\nu\ro_{10}}{(1+\nu)(\ro_{10}+\ro_{20}a^{-3\nu})}.
\ee
At early times and for $\nu>0$,  the effective energy density  behaves as $\ro\approx a^{-3}$, the effective barotropic index (\ref{eb}) $\ga\approx 1$ and  the effective fluid describes an Universe dominated by nearly pressureless dark matter. However, a late time accelerated Universe i.e., $\omega<-1/3$ with positive dark energy densities require that $\nu>1/2$, $\beta<1$ and $\al>1$. From now on we adopt the latter restrictions.


\section{Observational constraints:Hubble data vs. SNe Ia}
In what follows, we will provide a full estimation of the cosmological paramaters by constraining them with the Hubble data  \cite{obs3}- \cite{obs4} and the 557 SNe Ia data from the Union2 compilation \cite{Union2}. In the former case, the  statistical analysis is based on the $\chi^{2}$--function of the Hubble data which is constructed as (e.g.\cite{Press})
\be
\n{c1}
\chi^2(\theta) =\sum_{k=1}^{12}\frac{[H(\theta,z_k) - H_{obs}(z_k)]^2}{\sigma(z_k)^2},
\ee
where $\theta$ stands for cosmological parameters, $H_{obs}(z_k)$ is the observational $H(z)$ data at the redshift $z_k$, $\sigma(z_k)$ is the corresponding $1\sigma$ uncertainty, and the summation is over the 12 observational  $H(z)$ data. It should be stressed  that one of the main reason in using the Hubble data  is related to  the fact that the Hubble function is not integrated over. Further, the function $H(z)$  is directly related with the properties of the dark energy, since its value comes from the cosmological observations. Using the absolute ages of passively evolving galaxies observed at different redshifts, one obtains the differential ages $dz/dt$ and the function $H(z)$ can be measured through the relation $H(z)=-(1+z)^{-1}dz/dt$ \cite{obs3}, \cite{obs4}. The data  $H_{obs}(z_i)$ and $H_{obs}(z_k)$ are uncorrelated because they were obtained from the observations of galaxies at different redshifts. Since we are mostly interested in obtaining the bounds for the model parameters, we will adopt as prior  $H_{0}=72.2 \pm 3.6~{\rm km~s^{-1}\,Mpc^{-1}}$ \cite{H0} as it needed. The Hubble expansion of the model  becomes: 
\be
\n{Ht}
H(\theta; z)=H_0\Big\{B(1+z)^{3}+ (1-B)(1+z)^{3(\nu+1)}\Big\}^{\frac{1}{2(\nu+1)}}
\ee
\be
\n{B}
B[\theta]=\frac{\nu+1}{\nu}\left[\al(\Omega_{x0}-1) + (1-\beta\Omega_{x0}) \right]
\ee
where $\theta=\{\al, \beta, \Omega_{x0}, \nu\}$  and we have used that $\ro_{02}/\ro_{01}=(B-1)/B$. The two independent parameters $\al$ and $\beta$ will be fixed along the statistic analysis. Then, for a given pair of $(\al_{f}, \beta_{f})$,  we are going to perform the statistic analysis by minimizing the $\chi^2$ function to  obtain the best fit values of  the random variables $\theta_{c}=\{\nu, \Omega_{x0} \}$ that correspond to a maximum of Eq.(\ref{c1}). More precisely, the  best--fit parameters $\theta_{c}$ are those values where $\chi^2_{min}(\theta_{c})$ leads to the local minimum of the $\chi^2(\theta)$ distribution. If $\chi^2_{dof}=\chi^2_{min}(\theta_{c})/(N -n) \leq 1$ the fit is good and the data are consistent with the considered model $H(z;\theta)$. Here, $N$ is the number of data and $n$ is the number of parameters \cite{Press}. The variable $\chi^2$ is a random variable that depends on $N$ and its probability distribution is a $\chi^2$ distribution for $N-n$ degrees of freedom.

Besides, $68.3\%$ confidence  contours  in the $(\nu, \Omega_{x0} )$ plane  are made of the random data sets that satisfy the inequality $\Delta\chi^{2}=\chi^2(\theta)-\chi^{2}_{min}(\theta_{c})\leq 2.30$. The latter equation defines a bounded region by a closed area around $\theta_{c}$ in the two-dimensional parameter plane, thus the $1\sigma$ error bar can be identified with the distance from the $\theta_{c}$ point to the boundary of the  two-dimensional parameter plane. It can be shown that $95.4\%$ confidence contours  with a $2\sigma$  error bar in the samples satisfy $\Delta\chi^{2}\leq 6.17$ while the data within $99.73\%$ confidence  contours  with a $3\sigma$ error bar are accommodated in the domain defined by $\Delta\chi^{2}\leq 11.8$. After performing this analysis  we are in position to get  confidence  contours in the $(\nu, \Omega_{x0} )$ plane, thus using the  $\chi^2(\al_{f}, \beta_{f}, \nu, \Omega_{x0} $) distribution  one can find the $68.3\%$, $95.4\%$, and $99.73\%$ confidence  contours respectively. We have taken the point of reference $(\al_{f},\beta_{f})=(1.01,0.15)$ but it is possible to show a wide set of admissible values for $\al$ and $\beta$ which leads to a good fit [see Table(\ref{VP}) ]. Thus, from this analysis we get the best fit at $\theta_c=(\nu,\Omega_{x0})=(1.19 \pm 0.12; 0.61 \pm 0.02 )$. It corresponds to a local minimum $\chi^2_{min}=7.86$ leading to a good fit with $\chi^2_{dof}=0.786$ per degree of freedom. The Ricci's  cutoff $(\al, \beta)=(4/3,1)$ does not guarantee the convergence of the minimization process. However, the values of ($ \nu,\Omega_{x0}$) obtained from an holographic dark energy $\ro_{x} \propto R$, namely $(4/3,\beta)$, fulfills the goodness condition $\chi^2_{dof}<1$.  The values of $\Omega_{x0}$, which varies from, $0.58$ to $0.69$, do not deviate significantly from the observational limits provided by the WMAP-7 project \cite{WMAP7} with  $\Omega_{x0}=0.73$ [see Table (\ref{VP})]. Comparing the Ricci model with the one arising from MHRDE for $(\alpha=1.01,\beta=0.15)$, the former gives  $(\nu,\Omega_{x0})=( 1.19, 0.69)$, whereas the latter yields $(\nu,\Omega_{x0})=( 1.19, 0.61)$,  so the Ricci model seems to be statistically favored by $H(z)$ data  showing a $\Omega_{x0}$ closer to the observational bound reported by  the WMAP-7 project \cite{WMAP7}. 

\begin{figure}[h!]
\begin{center}
\includegraphics[height=6cm,width=7.5cm]{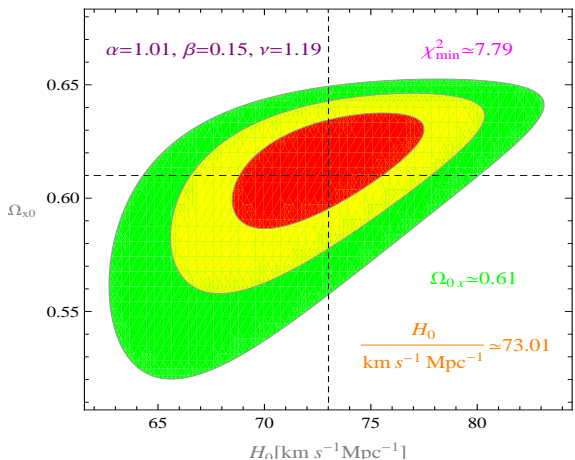}
\caption{ We show the confidence contour in the $H_{0}-\Omega_{x0}$  plane for the MHRDE model considering as priors $\nu=1.19$, $\al=1.01$ and $\beta=0.15$. The best fit value of  $(H_{0},\Omega_{x0})$ is used to compare with the $\Lambda$CDM model. }
\label{1b}
\end{center}
\end{figure}

We  estimate the best value of $H_{0}$ and $\Omega_{x0}$ for the $\Lambda$CDM model using the Hubble data as well as the Union2 data for SNe Ia \cite{Union2}.  The former dataset  leads to $H_{0}= 73.60 \pm 3.18 ~{\rm km~s^{-1}\,Mpc^{-1}}$ and $\Omega_{x0}= 0.730 \pm 0.04$ with $\chi^2_{dof}=0.770$ whereas the latter one gives $H_{0}= 70 ~{\rm km~s^{-1}\,Mpc^{-1}}$ and $\Omega_{x0}= 0.73 $ along with $\chi^2_{dof}=0.978$. Now, in oder to make possible a comparison with the $\Lambda$CDM model we need to estimate the same types of parameters so we take as priors $\nu=1.19$, $\al=1.01$, and $\beta=0.15$ but we allow $H_{0}$ and $\Omega_{x0}$ as free parameters to be found under the minimization process. It turns out that $H_{0}= 73.011 \pm 2.97 ~{\rm km~s^{-1}\,Mpc^{-1}}$ and  $\Omega_{x0}=0.617 \pm 0.001$ with $\chi^2_{dof}=0.779<1$ [see Fig.(\ref{1b})], then both models give cosmological bounds of the pair  $(H_{0}, \Omega_{0x})$ very consistent with the those reported in \cite{WMAP7}.

In order to compare  the Hubble data (12 points) with the Union2 compilation of 557 SNe--Ia \cite{Union2}  we proceed as follows; thus, we
took as priors $H_{0}=72.2~{\rm km~s^{-1}\,Mpc^{-1}}$, $\al=1.01$ and $\beta=0.15$  in both cosmological data. We found  the best--fit values of $\nu$ and $\Omega_{x0}$ for both sets, focusing on the existence of  some tighter constraints coming from the SNe Ia data. For the Hubble data we obtained $\al=1.19$ and $\Omega_{x0}=0.61$ with $\chi^2_{dof}=0.786$ whereas the SNe Ia data lead to $\nu=1.5$ and $\Omega_{x0}=0.70$  with $\chi^2_{dof}=0.812 <1$; in broad terms the tighter constraints seems to be found with the Hubble data. Of course, these results can vary according to the parameter regions taken into account in the minimization process.

\begin{table}[htbp!]
\begin{center}
\begin{tabular}{r|c|c}
\hline\hline
$(\al, \beta)$&$(\Omega_{0x}\pm \sigma, \nu \pm \sigma)$&$\chi^{2}_{dof}$\\
&$$&$$ \\
\hline\hline
(1.01,0.05)&$(0.58\pm 0.20, 1.19 \pm 1.13)$&0.786\\
(1.01,0.1)&$(0.58\pm 0.23, 1.19 \pm 1.13)$&0.786\\
(1.01, 0.15)&$(0.61 \pm 0.02, 1.19 \pm 1.13)$&0.786\\
(1.2,-0.1)&$(0.55 \pm 0.16, 1.19 \pm 1.13)$&0.786\\
(1.2,-0.05)&$(0.57 \pm 0.17, 1.19 \pm 1.13)$&0.786\\
(4/3,-0.1)&$(0.59 \pm 0.15, 1.19 \pm 1.13)$&0.786\\
(4/3,0.1)&$(0.69 \pm 0.17, 1.19 \pm 1.13)$&0.786\\
\hline\hline
\end{tabular}
 \end{center}
    \caption{\label{VP} We show the observational bounds for the pair $(\Omega_{x0}, \nu)$ varying $(\al,\beta)$ with a given prior of $H_{0}=72.2 \pm 3.6~{\rm km~s^{-1}\,Mpc^{-1}}$}
\end{table}

\begin{figure}
\begin{center}
\includegraphics[height=6cm,width=7.5cm]{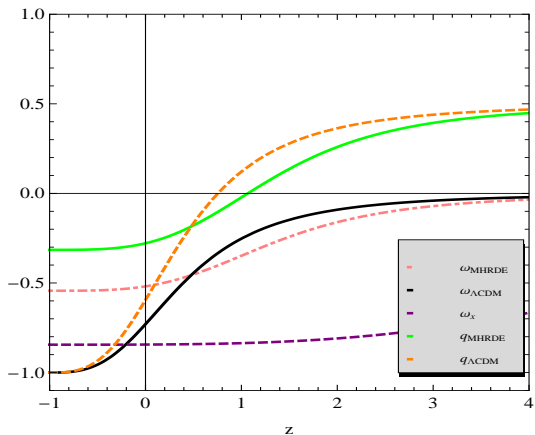}
\caption{We show the equations of state  for the effective fluid $\omega(z)$,  the dark energy $\omega_{x}(z)$, and the deceleration parameter  $q(z)$.}
\label{fig5}
\end{center}
\end{figure}

Now, using the best--fit model parameters $\theta_c=(\nu,\Omega_{x0})=(1.19 \pm 1.13, 0.61 \pm 0.02 )$ we would like to compare the   model having a MHRDE with the standard $\Lambda$CDM scheme composed of baryonic matter and a constant dark energy $\Omega_{0x}=0.73 \pm  0.04$. As $-1\leq\omega(z),\omega_{x}(z)\leq 0$, the equations of state of the effective fluid and dark energy do not cross the phantom line, at least for the best--fit model parameters used previously [see Fig.(\ref{fig5})], therefore this model does not exhibit a quintom phase \cite{quinto}.
 The expression of $\omega_{x}$ at present 
\be
\n{wx0}
\omega_{x0}= -\frac{\nu(\alpha -\beta)B}{(\alpha-1)(1+\nu)+\nu B}, 
\ee
becomes $\omega_{x0}=-0.88$ when evaluating at the best--fit values $\theta_c=(\nu,\Omega_{x0}, \alpha, \beta)=(1.19, 0.61, 1.01, 0.15)$. It is  close to the value reported by WMAP-7, $\omega_{x0}=-0.93$,  when the joint analysis of the WMAP+BAO+$H_{0}$+SN data \cite{WMAP7} for constraining the present-day value of the equation of state for dark energy is made.

Fig. (\ref{fig5}) shows the evolution of the decelerating parameter $q=-\ddot{a}/aH^{2}$ with the redshift $z$ for the MHRDE and $\Lambda$CDM models.  It takes the form 
\be
\n{q0}
q_{0}= \frac{1+ \nu(1-3B) }{2(1+\nu)}, 
\ee
at $z=0$ for the former model. Using the best--fit values $\theta_c=(\nu,\Omega_{x0}, \alpha, \beta)=(1.19, 0.61, 1.01, 0.15)$ in Eq. (\ref{q0}), one gets  $q_{0}=-0.27$ while for the $\Lambda$-CDM model one obtains $q_{0}=-0.59$. The critical redshift where the acceleration starts, 
\be
\n{za}
z_{acc}= -1+ \Big[\frac{(2\nu-1)B}{(1+\nu)(1-B)}\Big]^{1/3\nu}, 
\ee
turns to be $z_{acc}=1.06$ for the best--fit values $\theta_c$, then our model enters the accelerated regime earlier than the  $\Lambda$CDM one with $z_{acc}=0.75$.

\begin{figure}
\begin{center}
\includegraphics[height=6cm,width=7.5cm]{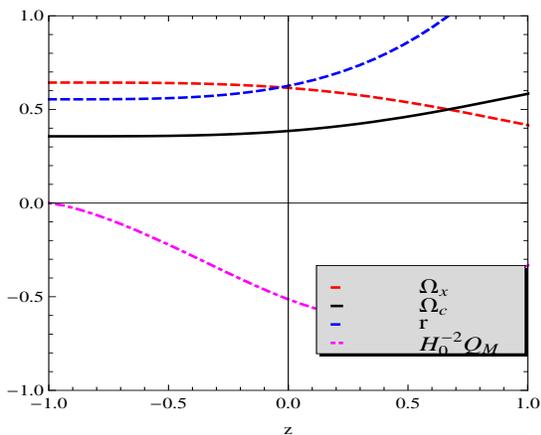}
\caption{The density parameters ($\Omega_x$, $\Omega_c$), the ratio $r=\Omega_c/\Omega_x$, and $H^{-2}_{0}Q$ are shown versus the redshift $z$.}
\label{fig7}
\end{center}
\end{figure}

In Fig.(\ref{fig7}) we plot the density parameters $\Omega_c$, $\Omega_x$, its ratio $r(z)$ and find the present-day values of $\Omega_{x0}=0.61$, $\Omega_{c0}=0.39$ and $r=0.62$. It shows that the  model with a MHRDE seems to be  appropriated for resolving the coincidence problem. Regarding the  modified Ricci coupling function,  one can show that $Q\leq 0$  and the coupling decreases its strength with the redshift and goes to zero in the far future, $z \rightarrow -1$ [see  Fig.(\ref{fig7})].


As a closing comment, we would like to address a discussion concerning the values of $\al$ and $\beta$ taken into account through this section. Here we have focused on the transition of the Universe between a stage dominated by dark matter followed by an  era  dominated by the holographic dark energy that  makes the Universe exhibt an accelerated expansion (present-day scenario) and, in both stages  a nonlinear interaction in the dark sector has been taken into account. In order to estimate the parameter $\nu$ and $\Omega_{x0}$ we have used the values of $\al$ and $\beta$ which are consistent with the $\chi^{2}$-statistical analysis because  they fulfill the condition $\chi^{2}_{dof}<1$. Now,   we are going to explore a modification on the aforesaid model  by adding a third component, say $\ro_{m}$, which does not interact with $\ro_{c}$ and $\ro_{x}$. The total energy density reads as  $\ro_{t}=\ro_{m}+ \ro$ with $\ro=\ro_{c}+\ro_{x}$ and  the MCE  is split  as
\be
\n{nc1}
\ro'+ \al\ro-(\al-\beta)\ro_{x}=0,
\ee
\be
\n{nc2}
\ro_{m}'+ \al\ro_{m}=0
\ee
so (\ref{nc1}) is equal (\ref{06})  while (\ref{nc2}) shows that the third component does not transfer energy to the interacting dark sector. From (\ref{nc2}) one finds that $\ro_{m}=\ro_{m0}a^{-3\al}$ whereas the behavior of dark matter and dark energy with the scale factor  can be obtained from  (\ref{cI}) and (\ref{xI}), respectively.   Using (\ref{cI}), (\ref{xI}) and (\ref{nc2}) one gets the Hubble parameter in term of the redshift $x=1+z$ and the relevant cosmological parameters

\be
\n{Ht2}
\frac{H(z)}{H_0}=\Big[(1-\Omega_{m0})\big(Bx^{3}+ (1-B)x^{3(\nu+1)}\big)^{\frac{1}{(\nu+1)}}+\Omega_{m0}x^{3\alpha}\Big]^{1/2}
\ee
\be
\n{B2}
B[\theta]=\frac{\nu+1}{\nu}\Big[\frac{\al-\beta}{1+\frac{1-\Omega_{x0}-\Omega_{m0}}{\Omega_{x0}}} + (1-\al)\Big]
\ee
where the flatness condition $1=\Omega_{x0}+\Omega_{c0}+\Omega_{m0}$ has been used. In what follows we  would to examine two  traits  of the model. First, we fix $\al=4/3$ to get a radiation contribution in the total density because it will address the problem of the dark energy at early times; thus,  as is well known the fraction of dark energy in the radiation era should fulfill the stringent bound $\Omega_{x}(z\simeq 1100)<0.1$ in order for the model be consistent with the big bang nucleosynthesis (BBN) data. For the priors $(H_{0}=72.2, \nu=1.19, \Omega_{x0}=0.7, \al=4/3)$ the Hubble data give as the best--fit values $\beta=0.1$ and $\Omega_{m0}=1.7\times 10^{6}$ along with $\chi^{2}_{dof}=0.78< 1$ with a fraction of dark energy $\Omega_{x}(z\simeq 1100)=0.2$ nearly close to the BBN's bound. Second, employing the Hubble data for (\ref{Ht2}) we estimate the best--fit value of $\Omega_{m0}$ and $\alpha$. Taking as priors  $(H_{0}=72.2, \nu=1.19, \Omega_{x0}=0.61, \beta=0.15)$ the $\chi^{2}$--analysis yields as the best--fit values $\al=1.01$ and $\Omega_{m0}=9.9\times 10^{-5}$ together with a $\chi^{2}_{dof}=0.79< 1$. Moreover, the latter case leads to an early  dark energy  $\Omega_{x}(z\simeq 1100)=0.01<0.1$ which is consistent with the bounds reported in \cite{EDE1} or with the future constraints achievable by Planck and CMBPol experiments \cite{EDE2}.  Therefore, taking the third component as the radiation term or a nearly radiation contribution, has helped to validate the first model,  indicating that the value of the cosmological parameters selected are consistent with BBN constraints. 

\section{conclusion}
We have explored an interacting dark sector with a MHRDE, where the IR cutoff  is provided by the modified Ricci scalar. We have introduced an interaction between the dark matter and dark energy densities, homogeneous of degree 1 in the variables $\ro$ and $\ro'$, and solved the source equation for the total energy density of the mix. Further, the equation of state of the effective fluid is that of the relaxed Chaplygin gas,  interpolating between a matter dominated phase at early times and an accelerated expanding phase dominated by the MHRDE at late times. 

We have used the observational Hubble data to constrain the cosmological parameters of the model and to compare with the $\Lambda$CDM model.  Taking as a reference point  $(\al_{f},\beta_{f})=(1.01, 0.15)$  we get the best fit at $\theta_c=(\nu,\Omega_{x0})=(1.19, 0.61)$ with $\chi^2_{min}=7.86$ leading to a good fit with $\chi^2_{dof}=0.786 <1$ per degree of freedom. We have established that a model with a holographic dark energy $\ro_{x} \propto R$ leads to $0.59<\Omega_{x0}<0.69$ which is close to the bounds $\Omega_{x0}=0.73$ provided by  WMAP-7 \cite{WMAP7}. For $\beta=0.1$, we have shown that the Ricci cutoff ($\alpha=4/3$) is  consistent with other values of $\alpha$ because  it fulfills the goodness condition ($\chi^2_{dof}<1$). In addition, we have obtained the allowed range of $(\nu,\Omega_{x0})$ when one varies $\al$ and $\beta$  [see Table.(\ref{VP})]. Properly estimating the $H_{0}$ and $\Omega_{x0}$ with the Hubble data we have confronted the $\Lambda$CDM  with  the MHRDE model; thus, both models give some bounds of the pair  $(H_{0}, \Omega_{x0})$ consistent the those reported in \cite{WMAP7}. Besides, we have taken into account the SNe Ia with the Union2 data for calculating the best--fit values of $\nu$ and $\Omega_{x0}$. It led to $\nu=1.5$ and $\Omega_{x0}=0.70$ with $\chi^2_{dof}=0.812 <1$  while the Hubble data gave $\nu=1.19$ and $\Omega_{x0}=0.61$ with  $\chi^2_{dof}=0.786$. 

We have found that the equations of state of the dark energy equation and the unified fluid, at the best--fit values $\theta_c$, do not cross the phantom divide line [see  Fig.(\ref{fig5})] while the present value of the equation of state for the dark energy is $\omega_{x0}=-0.88$.

From the deceleration parameter [see Fig.(\ref{fig5})] and the best fit values $\theta_c$, we have obtained that the acceleration starts at $z_{acc}=1.06$ hence, the model with a MHRDE enters the accelerated regime earlier than the $\Lambda$CDM with $z_{acc}=0.75$. We have shown that the density parameters $\Omega_c$, $\Omega_x$, and its ratio $r(z)$ in  Fig.(\ref{fig7}) seem to alleviate the coincidence problem. It is related to the decreasing behavior of the interaction with the redshift and its vanishing limit in the far future [see  Fig.(\ref{fig7})].

In order to examine if the value of the parameters obtained through the Hubble and/or SNe Ia data are consistent with the physic at primordial eras such recombination one ($z \simeq 1100$), we have included a non-interacting component  for studying the behavior of dark energy at early times. Interestingly enough, we have found that our model is consistent with the stringent bounds $\Omega_{x}(z \simeq 1100)<0.1-0.2$ reported in the literature, further it turned that the aforesaid model together with the cosmological constraints obtained with the Hubble data are in agreement with the future constraints achievable by Planck and CMBPol experiments \cite{EDE2}. 

\acknowledgments
We wish to thank CosmoSul's organizers for giving us the opportunity to present this talk and for the delightful time passed at Rio do Janeiro during the meeting. LPC thanks  the University of Buenos Aires  for their support under Project No. X044 and the Consejo Nacional de Investigaciones Cient\'{\i}ficas y T\' ecnicas (CONICET) through the research Project PIP 114-200801-00328. MGR is partially supported by CONICET.

\vspace{1cm}

\end{document}